\documentclass[journal]{IEEEtran} 
\usepackage{hyperref}
\usepackage{graphicx}
\usepackage{latex8}
\usepackage{times}

\begin{document}

\title{A Call-Graph Profiler for GNU Octave}

\author{
\authorblockN{ Muthiah Annamalai \\}
\authorblockA{ University of Texas, Arlington, USA,\\  
 email: muthiah.annamalai@uta.edu\\ \\}
\and
\authorblockN{ Leela Velusamy \\}
\authorblockA{Department of Computer Science \& Engineering,\\ 
National Institute of Technology,\\
 Tiruchirapalli, India. \\
 email: leela@nitt.edu\\
}
}

\maketitle
\thispagestyle{empty}

%
% This paper needs to discuss the b/g
% 1. What we have done, and demonstrated.
% 2. What are the nuts & bolts of the work (Algorithm etc)
% 3. Show outputs.
% 4. Conclusions.
%

%Remove it Later.
%\footnote{14th May 2007, This is an preprint version of the paper.}

\begin{abstract}
We report the design and implementation of a call-graph profiler for GNU Octave,
a numerical computing platform. GNU Octave simplifies matrix computation for
use in modeling or simulation. Our work provides a call-graph profiler, which is
an improvement on the flat profiler. We elaborate design constraints of building
a profiler for numerical computation, and benchmark the profiler by comparing it to
the rudimentary timer start-stop (tic-toc) measurements, for a similar set of programs.
The profiler code provides clean interfaces to internals of GNU Octave, for other (newer)
profiling tools on GNU Octave.
\end{abstract}

%
% SEC I
%
\section{Introduction}
GNU Octave is a numerical computing platform based on programming with matrices as a 
fundamental construct. Several modeling and simulation research work benefit through 
the reduced time and complexity offered by matrix programming environments. Modeling
and simulation applications when developed through exploratory analysis, or initial 
speculative analysis of parameters, optimization in not the key goal. However, once 
a stable model is created, running parametric search over a large search space requires
the creation of optimized models.

Optimizing computational models written in matrix languages, for time and space usage,
help to directly speed up the search process by reducing time and resources (memory, diskspace).
There are many ways of optimizing program performance, like tracing, logging, and profiling.
Profiling is the only method where the runtime is proportional to program size, unlike the
other methods which depend on the execution time of the program. 

Clearly, for large scale parametric searches, we cannot use tracing or 
logging for optimization, due to enormous log data that needs post-analysis. 
This makes profiling as an efficient tool for program optimization.
Profiling for matrix programming languages like GNU Octave, 
help identify resource usage, inefficient function usage, program flow, piecemeal 
function run-times and a complete idea of the running times spent on subroutines as a 
percentage of the program lifetime.

The landmark work in profiling is the GNU Debugger (GDB) \cite{GDB}, which initiated the idea of a
Call-Graph profiler. Since GDB, many types of profilers  have been proposed 
and built, including dynamic instrumented profilers, static sampling profilers, flat profilers,
call graph profilers.

 The Java Virtual Machine (JVM) provides a complete infrastructure for
dynamic program analysis, and encourages custom built profilers for querying and collecting
statistics from the JVM \cite{JVMPI}. This is called the Java Virtual Machine profiling Interface (JVMPI),
and represents the state-of-the art in profiling. Using JVMPI many successful profilers can be built for generic
profiling or integrated into the existing applications.

We present a call-graph profiler for matrix based languages, implemented on the 
freely available (Open Source) GNU Octave platform. With knowledge of Octave 
internals \cite{OCTAVEGTK},  we chose GNU Octave platform
for creating profiler. Our work addresses concerns of optimization of numerical computational models, 
discrete simulations, and exploratory analysis, through the use of profiling.
 Important metrics for the profiler are minimal sampling-time, minimal memory
 and resource usage from profiler, dynamic data collection, meaningful output presentation.
 We build a dynamic instrumentation profiler, with dispatch built 
into the Octave interpreter, which allows us to create profilers of
increasing complexity from Flat profiler till the Call-Graph profiler.

The terminology in use while describing the profiler statistics are described as follows
\begin{enumerate}
\item \textit{Total time}: time taken for the subroutine to run, excluding the runtime of subroutines it calls.
This includes only the computational times, and not the times for functions called during computation.
\item \textit{Self time}: the run-time of a subroutine including its calls to the sub-functions. This is greater
than or equal to the total-time of the subroutine.
\item \textit{Average time}:  the average of the total times, over all calls to the subroutine. The self time,
and total time are reported as averages over the total calls.
\item \textit{Percentage time}:  the fraction of the program runtime for which the given subroutine has
been executing.
\item \textit{Number of calls}: total number of times a subroutine was called.
\item \textit{Cumulative time}: This gives the sum of the times of all subroutines, that have lesser self time
than the current function. Results are sorted in descending order of self time, and increasing cumulative time.
\end{enumerate}
In this paper, we use the terms function and subroutine interchangeably.

The rest of the paper is organized as follows. In Sec II, we report the profiling API required
to create profilers on GNU Octave. The algorithms used for the Flat profiler and the 
Call-Graph profiler are discussed in Sec III, IV. Benchmarking results of the profilers are 
presented in Sec V, to illustrate accuracy and confidence on the profiled results. Finally, 
in Sec VI,  we review the features and limitations of our profiler framework for Octave, 
and indicate the future work required to integrate the profiler into the main GNU Octave project.

%\cite{}

%
% SEC II
%
\section{Octave Profiler API}
%
% This is -NOT- User level API for the profiling tools 
% Profiler tools 
% API used
%
Most profiling systems work by collecting data  of a program at runtime. Profilers
collect statistics, and the time of occurrence for each piece of information. To enable
collection of profiling events like function invocation, object creation, deletion, 
function call completion, and such interpreter-system specific data, profilers need 
an event delivery mechanism from the interpreter itself. This makes building a profiler
for GNU Octave language as two tasks; building interpreter API event delivery mechanism
to the profiler, and then building profilers to make use of the events. This separation of
concerns, was inspired from the profiler design of languages like Python \cite{PYTHON},
 and Ruby \cite{RUBY}. The rest of this section describes the events of interest to our	
profiler, and API for the event delivery system.

The simplest profiler  (Flat profilers, Sec III), need information about the occurrence 
of two events; the function call, and function return. Since Octave is not an Object oriented
system we do not have Object creation, deletion, nor member function invocation and such events
cannot be reported.In a more complicated profiler (Call-Graph profiler, Sec IV), the same 
events (function call, return) have attributes like the function-caller and function-callee
passed to the profiler program. We do not trap events like variable state changed, entering a
program section corresponding to line number in source code, or leaving such a program section, etc.
The implications of these events are discussed in Sec VI.

The API design of the event-reporting system is based on a global singleton class \textit{octave\_profiler},
which is instantiated by the Octave interpreter. The class \textit{octave\_profiler}, has a 
member variable \textit{profiler\_fcn} which serves as a event reporter function. This is
set by the profiler program, using a call to \textit{static bool set\_profiler(profiler\_function profile);}.
When set by a profiler, the \textit{profiler\_fcn} is invoked from within the Octave interpreter whenever 
execution invokes a function or returns from a function. The profiler stops profiling, by invoking the
\textit{static bool clear\_profiler();} which returns the handle \textit{profiler\_fcn} to a null, and 
prevents event delivery. 

The event delivery is managed by function   \\
\textit{static void send\_event (const octave\_function \*fcn, profiler\_function\_type ftype, profiler\_call\_state cstate);}\\ 
from the internals of the Octave interpreter. In this invocation \textit{fcn} is the function in question, and \textit{cstate} indicates execution has entered or returned from a function. 

This \textit{send\_event} function abstracts the call sequence \textit{profile\_fcn(fcn,ftype,cstate);} by keeping it private.

This API in short provides a hook for the profilers to start/stop receiving events, and a base class that implements this mechanism. Function profilers are written by deriving from the primitive base class \textit{profile\_base}. The base class provides timing information, and the setup tear-down for the events as discussed above.

The user level profiler interface is through the function \textit{profile}, which
has usage : \textbf{ profile [on\vline off\vline info] [graph\vline flat]}, 
where the options are indicated in square brackets. A typical use case would be: 
\begin{enumerate}
\item \label{profile_usage} profile on graph \textit{\%use a call-graph profiler} 
\item bch() \textit{\%invoke the script}
\item profile info \textit{\%stop profiling \& printout the results}.
\end{enumerate}

The \textit{profile} is responsible for creating instances of Call-Graph or Flat profilers, and pass on the start, stop, print requests to these profilers using API of the profiler, as 
\begin{enumerate}
\item \textit{static void start\_profiling()}
\item \textit{static void stop\_profiling()}
\item \textit{void print\_profile()}
\end{enumerate}
The  profilers to extend the base class \textit{profile\_base}. This completes the description of the Octave side profiling API, and the user interface for profilers. 

%
% SEC III
%
\section{Flat-File Profiler}
Flat-file profilers are very simple, and only count the average statistics of the program. The only attribute information saved by Flat-profilers are saved in the structure, \textit{call\_elem}, which has fields \\
\begin{enumerate}
\item  long int ncalls;
\item  double total\_time;
\item  double self\_time;
\item  std::string key;
\end{enumerate}

for each function that is invoked in the program. The meaning of each field, is self explanatory and the terms are defined in the introduction. Nowhere are absolute times measured, and programs rely on relative time separation of events. The timing information is stored in the class objects of \textit{time\_elem}.\\ 
\begin{enumerate}
\item  double delta; //incremental time from the previous routine that a function is called
\item double tick; //time for running our kids/child functions
\end{enumerate}
The relative times of events noted down at each event, are finally added up to obtain the statistics for Flat-profiler output.

\subsection{Implementation}
%
% 1. Algorithm Used.
% 2. Source API / Design Pattern (template)
% 3. User level API for the profiling tools 
%
The class \textit{profiler\_flat} implements the Flat-profiler, by extending \textit{profile\_base}. The algorithm is implemented using a stack data structure. The essential algorithm is the same for the complex Call-Graph profiler too.

The profiler dispatch function delivers the function invoke and return events by interfacing with the profiler API, describe in Sec II. The profiler
function for the Flat-profilers is given as,\\
\textit{static void profile\_func(const octave\_function *fcn,  profiler\_function\_type ftype,  profiler\_call\_state cstate);}

This function further delivers the events to the particular call-processing routines that handle event call, return separately. The  function \textit{profile\_func()} is modeled as a template pattern, that delegates the events to particular handlers.

\subsection{Algorithm}
The algorithm for the Flat-profiler is summarized as,
\begin{enumerate}
\item When profiler is started, note starting time.
\item Register the profile event handler
\item On Call Event:\\ Push a \textit{time\_elem} instance set to zero, into time-stack.This time-stack is a false call-stack, as it mirrors the interpreter call-stack, functions are invoked and returned.
\item On Return Event: 
\begin{enumerate}
\item Check if hashtable has an instance of record for the given function. Otherwise create a new \textit{call\_elem} instance for this function and set the name to the function.
\item Increase the number of calls on this record by 1.
\item Compute the relative time difference between the call and return events; Use the time\_elem object on the top of time-stack.
\item Add the total time to the call record's corresponding field.
\item Add the self time to the call record's corresponding field.Compute self time by subtracting from total time, the value of tick.
\item Update the record in the hashtable, indexed by the function name as key.
\item If the call-stack of time, is not empty add the cost of this call, to the parent in the parents, time element tick field.
\end{enumerate}
\item Repeat the steps 3-4, till stopped.
\item Clear the profiling handler, and receive no more events.
\item Once profiling is stopped, prepare to print output. Compute \% times.
\item Sort the hashtable entries according to the total-time field of record.
\item Print out according to descending order of total times.
\end{enumerate}

It is important to note the source of this algorithm is obtained from profilers for popular programming languages \cite{PYTHON}, \cite{RUBY}. We attribute the idea to the Python, and Ruby implementations.

%
% SEC IV
%
\section{Call-Graph Profiler}
Call-Graph profiler builds the profiling output with the program execution,as a directed graph with arcs. The arcs point from the caller to the callee, and conveys the time of execution of the callee function. Second order statistics and more than averages can be obtained by sifting through the profiling data, and it becomes much valuable than Flat profiling.

There are particular cases where Flat profiling information is not helpful; in general numerical routine execution times depend on the size of the input argument, and the average total time used for routines that are not O(1), skew the profiled data. Call-Graph profilers side step such problems by assigning second order statistics, which include the self , average, and total times for each arc of a function call, and profiled function's 
complexity can be clearly observed without skewing the data. From definition of a Call-Graph, the parent-child relationships (caller-callee relations) from the profiled information are also immediately available.

It is to be noted, that in our implementation not every parent-child relationship is saved,and the data is averaged for each unique caller-callee information, in order to reduce the profiler output to a meaningful subset.

Data structures derived from \textit{call\_elem}, and \textit{time\_elem} with extra variables,to contain the caller-callee relationship records are used.

\subsection{Implementation}
%
% 1. Algorithm Used.
% 2. Source API / Design Pattern (template)
% 3. User level API for the profiling tools 
%
The Call-Graph profiler is implemented in the 
\textit{class profile\_callgraph} which as in the Flat-profiler derives from the \textit{profile\_base} class.

The Call-Graph profiler is itself, so to speak, an incremental improvement over the Flat-profiler. It's profiler event reported, dispatch and logging mechanisms are similar to Flat-profiler, and not reiterated here.

\subsubsection{Algorithm}
Much of the algorithm of the call-graph profiler is very-similar to the Flat-profiler. The differences remain;
\begin{enumerate}
\item Function call-event: when call-stack is empty, all call events are added to the toplevel callee hashtable. This saves the caller-callee information.
\item Function return-event: the returning function is added as a callee to top-of-stack (TOS).Then the caller for this returning function set in the hash-table, and its timing record updated. Similarly the callee for the TOS function is set as the returning function, and the caller records updated.\item Printing: the data is printed out as a tree, after sorting according to descending order.
\end{enumerate}

The printing of results follows a tree like pattern, illustrating the Call-Graph nature of the program execution.

%
% SEC V
%
\section{Results \& Discussion}
To evaluate the Flat and Call-Graph profilers a test case comprising of a communication system simulation program was evaluated. The program, and associated files were about 1672 lines of Octave code, excluding comments. This code set is chosen for its availability as much as its similar performance on the Flat-profiler and the Call-Graph profilers, due to the constant input modulation sizes used all over the simulation program.

The profiling is carried on at the toplevel program using the sequence of calls to \textit{profile} function mentioned in Sec \ref{profile_usage}.

\subsection{Flat profiler results}
The Flat profiler, gives an average performance of the functions across the runtime of the program. The run time of the programs are reported in seconds, while the ms/call indicates milliseconds/call. It should be noted that measured results are more finely-granular than the ones reported. Results are rounded-off due to formatting constraints. 

From the results in Table 1, we see that CPU hogging function is GF\_add which takes about ~ 29\% of the program runtime. This information, along with the self-times and number of calls can be used to arrive at possible optimization candidates.

%put the table at the end.
\begin{table}[e]
\caption{Flat-profiler results}
\begin{tabular}{|l|l|l|l|l|l|l|}\hline
    \%  & cumulative &   self  &        &      self  &    total & function  \\
time &  time seconds & seconds & calls & ms/call & ms/call & name \\  \hline
0.87 & 1.58 & 1.58 & 231 & 6.84 & 8.05 & bpskmod \\  \hline
0.00 & 1.59 & 0.01 & 524 & 0.01 & 0.01 & zeros \\  \hline
0.01 & 1.60 & 0.01 & 1 & 14.21 & 300.46 & BCH\_setup \\  \hline
0.02 & 1.64 & 0.04 & 1 & 37.75 & 568.93 & BCH\_poly \\  \hline
0.03 & 1.70 & 0.06 & 1 & 60.90 & 286.25 & GF\_table \\  \hline
0.00 & 1.70 & 0.00 & 63 & 0.01 & 0.01 & eye \\  \hline
0.03 & 1.76 & 0.06 & 56 & 1.12 & 2.99 & GF\_convolve \\  \hline
0.51 & 2.68 & 0.92 & 106641 & 0.01 & 0.01 & length \\  \hline
1.90 & 6.11 & 3.43 & 396249 & 0.01 & 0.01 & size\_equal \\  \hline
3.33 & 12.12 & 6.01 & 792498 & 0.01 & 0.01 & isreal \\  \hline
1.90 & 15.57 & 3.44 & 396249 & 0.01 & 0.01 & all \\  \hline
0.01 & 15.59 & 0.02 & 278 & 0.07 & 0.08 & isscalar \\  \hline
0.00 & 15.59 & 0.00 & 278 & 0.01 & 0.01 & size \\  \hline
24.49 & 59.88 & 44.29 & 396249 & 0.11 & 0.14 & mod \\  \hline
28.91 & 112.15 & 52.27 & 543132 & 0.10 & 0.14 & GF\_add \\  \hline
0.01 & 112.18 & 0.03 & 5 & 5.37 & 26.94 & GF\_minimalpoly \\  \hline
10.24 & 130.69 & 18.51 & 387685 & 0.05 & 0.10 & GF\_mul \\  \hline
0.05 & 130.77 & 0.08 & 7008 & 0.01 & 0.01 & ones \\  \hline
12.79 & 153.91 & 23.14 & 3391 & 6.82 & 22.08 & GF\_product \\  \hline
0.01 & 153.92 & 0.01 & 210 & 0.04 & 0.04 & GF\_polarize \\  \hline
0.01 & 153.94 & 0.02 & 1 & 16.93 & 16.93 & save \\  \hline
0.01 & 153.95 & 0.01 & 49 & 0.19 & 0.19 & disp \\  \hline
0.01 & 153.96 & 0.02 & 462 & 0.04 & 0.04 & randn \\  \hline
0.15 & 154.24 & 0.27 & 20790 & 0.01 & 0.01 & linspace \\  \hline
0.20 & 154.60 & 0.37 & 210 & 1.74 & 1.81 & boxmuller \\  \hline
2.30 & 158.75 & 4.15 & 231 & 17.97 & 18.69 & bpskdemod \\  \hline
0.09 & 158.91 & 0.16 & 20790 & 0.01 & 0.01 & sign \\  \hline
0.45 & 159.72 & 0.81 & 210 & 3.84 & 584.58 & BCH\_decode\_berlekamp \\  \hline
11.37 & 180.29 & 20.57 & 15330 & 1.34 & 6.32 & GF\_polyeval \\  \hline
0.00 & 180.29 & 0.00 & 210 & 0.01 & 0.01 & isempty \\  \hline
0.01 & 180.31 & 0.02 & 441 & 0.04 & 0.04 & GF\_complement \\  \hline
0.11 & 180.50 & 0.20 & 210 & 0.93 & 140.21 & GF\_roots \\  \hline
0.00 & 180.50 & 0.00 & 1 & 0.00 & 180833.00 & \#toplevel \\  \hline
\end{tabular}
\end{table}

The Call-Graph profiler output is more involved as shown in Fig 1. The important difference is, the call times, count information are collected as the caller-callee basis, and reported so. The Call-Graph profiler in this case helps to identify functions that perform on a O($n^2$) complexity basis.

The voluminous output of the Call-Graph profiler is reduced to the first few lines for brevity sake, is presented in the Figure 1. The same benchmark program was run on the Call-Graph profiler as well.

\label{fig_ncalls)A}
\begin{figure}
\centering
\includegraphics[height=2.2in,width=3.5in]{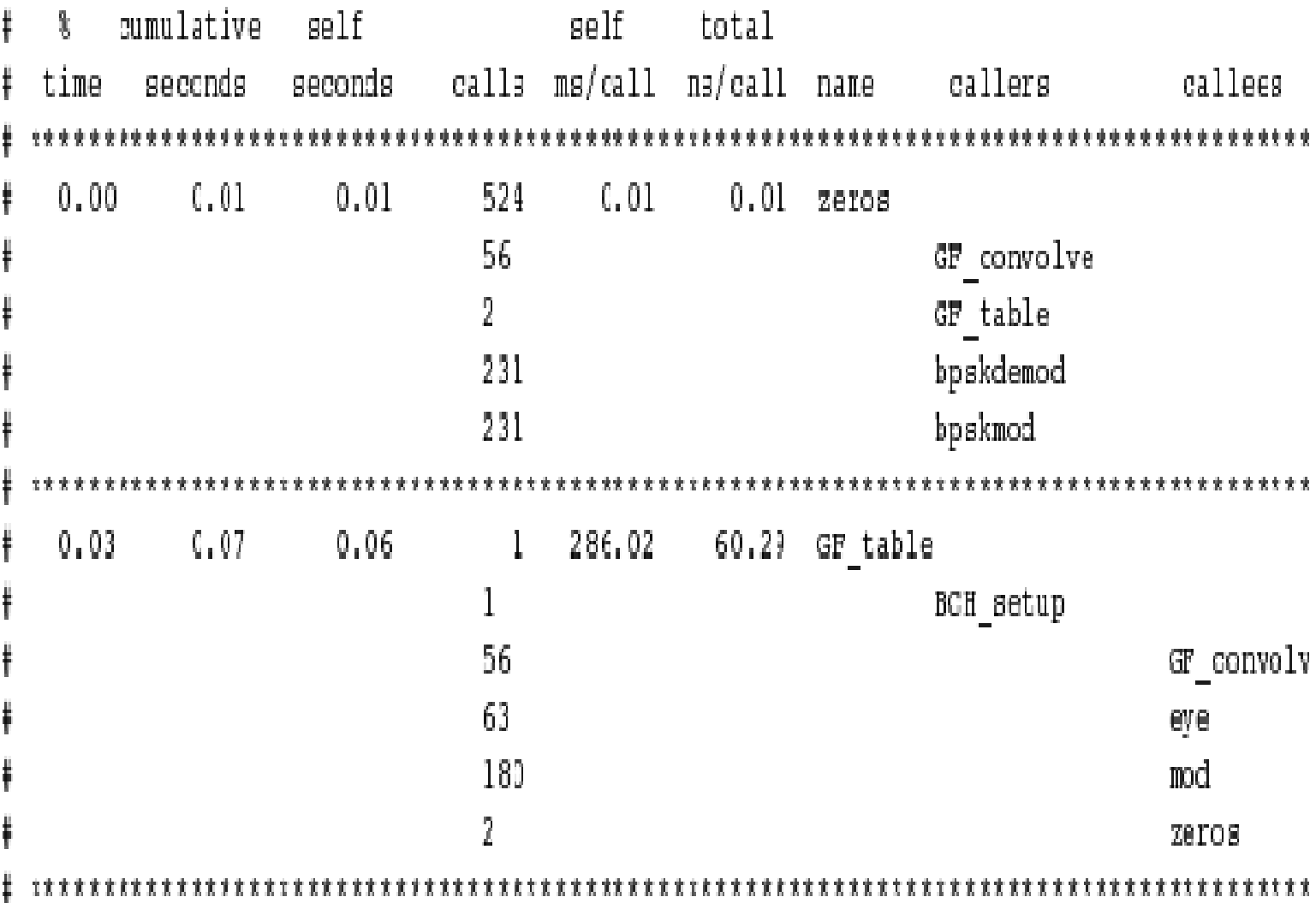} 
\caption{Call-Graph profiler output}
\end{figure}

\subsection{Profiling Overhead}
There is a significant performance hit due to the profiling. In our design, we explicitly compensate for the profiler runtime, and this is not a problem. The reported overhead times are found after compensation, and for the Flat-profiler. 

The reported overhead time can only be accounted for, using a free parameter  computed before profiling on each profiling session. This is called the \textit{bias} value, as reported in the Python profiler \cite{PYTHON}. In our profiler design we do not include such free parameters.

Such an overhead observed can only be attributed to the times that are not computable within the profiler. Our hypothesis attributes the time due to the interpreter's delay in invoking the profiler for each function call and return events. This agrees well with the observed O(n) overhead time dependence on the number of function calls. In Figure 2, a tight-loop function was profiled with a number of calls, to obtain the overhead information presented in the graph. The linear trend observed in the overhead time seems to justify the apparent constant overhead time for the interpreter which cannot be compensated without computing a constant bias factors. 

From Figure 2, the bias factor would seem to be the slope of the overhead time, which can be estimated to be around an overhead time of  8.1706x$10^{-06}$ seconds/call. This is however a free parameter, and dependent on implementation details. It should be noted that, most profilers suffer from the performance hit due to the profiling overhead.

On a more general note, from the benchmark tests we observe the total overhead times to be less than 0.5\% of the total program runtime. Our design has optimized the overhead time compared consistently from our initial prototype by compensating for each measurable profiling time.

The Call-Graph profiler performs with a larger overhead compared to the Flat-profiler; we estimate a rough factor of $2\times$ increase in the overhead time, for the Call-Graph profiler. The explanation for this variation we think, to be the memory creation and cleanup associated with the data structures used to build the Call-Graph. Also a non-trivial I/O times are associated with the Call-Graph procedures.

\label{fig_ncalls_B}
\begin{figure}
\includegraphics[height=2.5in,width=3.0in]{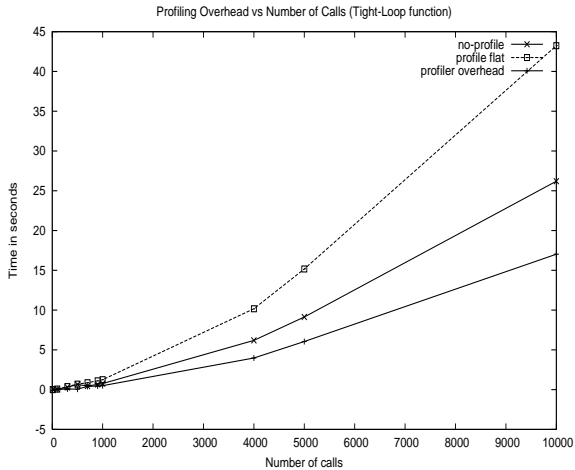}
\caption{Overhead time vs Number of Calls (Flat Profiler)}
\end{figure}

\subsection{Limitations of the profiler}
Certain features which are not implemented at present are not limitations to the profiler. These include 
\begin{enumerate}
\item resource profiling for opened-files, network-connections, database handles;
\item arguments passed from the caller-callee function are not traced;
\item  event filtering, and selective profiling.
\end{enumerate}

The limitations of the profiler reported below include features that cannot be added in the current design.
\begin{enumerate}
\item Memory profiling needs deeper access to the interpreter than the present framework can allow.
\item Line stepping and watch on variables are not possible, and more appropriate for a  debugger.
\item Non-local exits are not traced; This means uncaught exceptions are not profiled, and would end in a aborting of execution. This is however classified as a bug in the user's Octave script program.
\end{enumerate}

We also note that, complete integration of the Profiler into the codebase of the GNU Octave project requires a different approach to creating the Profiler-API. Such an profiler mechanism would work by walking the Abstract Syntax Tree (AST), and passing function call and return events to the profiler. From our experimental work, we see this as feasibility to bring the advantages of the call-graph profiler to Octave, in the future.

%
% SEC VI
%
\section{Conclusion}
%
%Rehash feature set, and status quo of project.
%
In this paper we have demonstrated a Flat profiler and a Call-Graph profiler for matrix based programming language like GNU Octave. We have reported the
profiling overhead, benchmark the performance for both the profiler. Further the limitations and possible extensions on this design are enumerated.
%\center{\textbf{Acknowledgments}\\}
%Thanks to John Eaton, David Bateman\\
% \& Paul Kienzle, from the GNU Octave Project.
\bibliographystyle{abbrv}

\end{document}